 \documentstyle[12pt]{article}
   \def\lsim{\stackrel{<}{\sim}}
   \def\beq{\begin{equation}}
   \def\eeq{\end{equation}}
   \def\ba{\begin{array}}
   \def\ea{\end{array}}

   \def\deg{\ifmmode{^{\circ}}\else ${^{\circ}}$\fi}
  
   \def\lsim{\,\raisebox{-0.13cm}{$\stackrel{\textstyle<}{\textstyle\sim}$}\,}
   \def\bi{\begin{itemize}}
   \def\ei{\end{itemize}}
   \def\bea{\begin{eqnarray}}
   \def\eea{\end{eqnarray}}
   \def\beas{\begin{eqnarray*}}
   \def\eeas{\end{eqnarray*}}

   \def\half{\frac{1}{2}}

   \def\dms{\delta m^2}
   
   \def\dplus{\Delta^{\rm m}_+}
   \def\dminus{\Delta^{\rm m}_-}
   
\begin{document}
   \begin{flushright}
   FERMILAB-Pub-00/300-T\\
   VAND--TH--00--8\\
   hep-ph/0011247\\
   November 20, 2000
   \end{flushright}
   \begin{centering}
  {\Large Optimizing T-violating Effects for Neutrino Oscillations
in Matter}\\
   \vspace{1cm}
   {\large Stephen J. Parke}$^1$ {\large and Thomas J. Weiler}$^2$\\
   \vspace{0.5cm}
   {\it $^1$ Theoretical Physics Department,\\
             Fermi National Accelerator Laboratory,
             Batavia, IL 60510, U.S.A.\\
  $^2$ Department of Physics \& Astronomy, Vanderbilt University,\\
   Nashville, TN 37235, U.S.A.}\\
   \end{centering}
   \vspace{1cm}
\begin{abstract}
\noindent
Matter effects may strongly enhance
the Jarlskog factor $J$ in T and CP-violating three-neutrino oscillation
probabilities.  
However, we show that when $J$ is enhanced, the same matter effects
{\it suppress} the oscillating factors
and {\it increase} the oscillation length.  The net result is
that there is no large enhancement in measurable probabilities
for earth-bound experiments using neutrino 
parameters suggested by current experiments.
We show that by an appropriate choice of the experimental 
parameters, neutrino energy and travel length, 
the T-violating probability can be enhanced
by matter effects over their vacuum values by 50\%.
Our approach is analytical, allowing considerable insight
into the underlying physics.
\end{abstract}

\section{Introduction}
Analogous to the CKM mixing-matrix in the quark sector of the
Standard Model (SM), there is an MNS mixing-matrix \cite{MNS}
in the lepton sector.  For three light neutrinos 
(assumed throughout this work) the  MNS matrix consists
of three angles and one (three) phase(s) for Dirac (Majorana)
neutrinos.  Regardless of the nature of the neutrinos, Dirac
vs.\ Majorana, neutrino oscillation experiments are sensitive
to a single phase through the measurement of a T-violating
asymmetry 
$P(\nu_\alpha\rightarrow\nu_\beta)
   -P(\nu_\beta\rightarrow\nu_\alpha)$
or a CP-violating asymmetry
$P(\nu_\alpha\rightarrow\nu_\beta)
   -P(\bar{\nu}_\alpha\rightarrow\bar{\nu}_\beta)$.
Much progress has been made toward determining the values of
the three mixing angles.
From measurements of the neutrino survival probabilities
$\nu_\mu\rightarrow \nu_\mu$ and $\nu_e\rightarrow \nu_e$
in the atmospheric flux, one infers that one mixing-angle 
is near maximal ($\pi/4$), and one is small \cite{atm}, 
the latter statement 
supported also by the Chooz experiment \cite{CHOOZ}.
From the $\nu_e\rightarrow \nu_e$ survival probability
in the solar flux, 
one infers that the third angle is either large (for the 
large-angle MSW (LAM) and the long-wavelength vacuum (LWV) 
solar solutions), or very small 
(for the small-angle MSW (SAM) solar solution) \cite{solar}.
Nothing is known about the T and CP-violating phase.

With construction underway for long-baseline terrestrial
oscillation experiments, attention has turned toward more precise 
measurements of the MNS parameters, including the phase.
Measurement of the T and CP-violating asymmetries appears
impossible for the SAM and LWV solutions.
For the SAM solution, this is because of the smallness of 
two of the three angles, while 
for the LWV solution, this is because of the extreme hierarchy 
of mass-squared values 
$\delta m^2_{\rm sun}/\delta m^2_{\rm atm}\sim 10^{-7}{\rm eV}^2$.
On the other hand, for the LAM solution, 
measuring T and CP-violating asymmetries to determine
the phase $\delta$ is more promising, with effects at the per cent level.
In terrestrial experiments, the neutrino beam will travel underground,
and one may ask whether earth-matter effects \cite{matter} can enhance 
the T and CP-violating probabilities \cite{3numatter}.

Throughout this paper, we assume a three neutrino world,
and the vacuum hierarchy $\delta m^2_{21}\ll \delta m^2_{32}$
as indicated by experiment.
We focus on T-violation, defined by the asymmetry
\beq
P^{\not {\rm T}}\equiv 
P(\nu_\alpha\rightarrow\nu_\beta)
   -P(\nu_\beta\rightarrow\nu_\alpha)
\,.
\label{Tasym}
\eeq
Here $\alpha$ and $\beta$ denote {\it different} 
neutrino (or anti-neutrino) flavors.
The CP-violating asymmetry, defined by
\beq
P^{\not {\rm C}\not {\rm P}}\equiv
P(\nu_\alpha\rightarrow\nu_\beta)
   -P(\bar{\nu}_\alpha\rightarrow\bar{\nu}_\beta)
\,,
\label{CPasym}
\eeq
contains the same contribution as $P^{\not {\rm T}}$ but includes additional
contributions arising solely from matter effects.
We do not pursue this complication in this work.
Here we analyze the case for a measurable T-violation.
Time-reversing the path of the neutrino through the earth gives
no extrinsic T-violation (for a spherically symmetric earth-matter
distribution).  This fact, and the relatively small effects of matter
on $P^{\not T}$ derived herein, make the T-violation measurement
an attractive approach for extracting the intrinsic T and CP- violating
parameter $\delta$.

For certain small values
of mixing angles and for certain neutrino energies,
strong enhancements may occur for the Jarlskog
invariant $J$ in matter (m) relative to vacuum (v) \cite{YKT00}.
It is not hard to see why $J_{\rm m}$ can be strongly enhanced near
a small-angle MSW resonance.
In the two flavor approximation, which is valid for three neutrinos
with a hierarchy of vacuum mass-squared differences
$\delta m^2_{21}\ll \delta m^2_{32}$,
the relation
$\delta m^2_{\rm m} =\sin (2\theta_{\rm v})\,\delta m^2_{\rm v}$
obtained at resonance,
implies $J_{\rm m}\propto 1/\theta_{\rm v}$
for small mixing-angle.
The purpose of this Letter is to
show that although $J_{\rm m}$ may be strongly enhanced
by small-angle matter resonances,
the measurable T-violating neutrino oscillation probability
proportional to $J_{\rm m}$ is not enhanced 
for earth-bound experiments.
In fact, depending on the
exact values of the mixing angles and vacuum masses, 
matter effects, if important at all, generally suppress these probabilities.
However, we do show that for an appropriate choice of neutrino
energy and travel distance, matter effects give modest 
($\sim$ 50\%) enhancements
of the T-violating probabilities for {\it terrestrial} experiments.
Our results are mainly analytical,
and so some insight into the physics of the matter phenomenon emerges.

\section{Enhancement of T-Violating Probabilities?}
The reason for the absence of a significant enhancement in probability
is easy to understand, as we now demonstrate.
For oscillations in vacuum the T-violating probability,
eq.\ (\ref{Tasym}), is simply
\beq
P^{\not {\rm T}}_{\rm v}= 16 J_{\rm v}
 \sin\Delta^{\rm v}_{21} \sin\Delta^{\rm v}_{32} \sin\Delta^{\rm v}_{31}\,,
\label{Tvac}
\eeq
where
\beq
\Delta^{\rm v}_{jk} = 
\frac{\delta m^2_{jk}|_{\rm v} L}{4E_\nu} 
= 1.2669\cdots\;\frac{(L/10^3 {\rm km})
(\delta m^2_{jk}/10^{-3}{\rm eV}^2)}{(E/{\rm GeV})}\,;
\label{Delta}
\eeq
$\delta m^2_{jk}|_{\rm v}$ is the difference of 
$j^{\rm th}$ and $k^{\rm th}$
vacuum mass-squared eigenvalues, $E_\nu$ is the neutrino energy,
and $L$ is the travel distance.
The Jarlskog factor \cite{Jarlskog}, $J$, 
in the standard mixing parameterization 
\cite{PDG} is given by
\beq
J_{\rm v}=
\left[ s_{21}\,s_{31}\,s_{32}\,c_{21}\,c^2_{31}\,c_{32}\,
\sin\delta\right]_{\rm v}
\label{Jvac}
\eeq
where $s_{21}\equiv \sin\theta_{21}$, etc.
$J_{\rm v}$ has a maximum value
of $\frac{1}{6\sqrt{3}}$.
Whereas for oscillations in matter of constant density\footnote
{In the earth this approximation is quite accurate  
for those paths that do not enter the earth's core; for 
paths that do enter
the core, the constant density approximation gives 
qualitatively the correct physics.} 
we have
\beq
P^{\not {\rm T}}_{\rm m}= 16 J_{\rm m}
 \sin\Delta^{\rm m}_{21} \sin\Delta^{\rm m}_{32} \sin\Delta^{\rm m}_{31}\,,
\label{Tmat}
\eeq
where the m sub- or superscript indicates the value in matter. 
Again the maximum value of 
$J_m =
\left[ s_{21}\,s_{31}\,s_{32}\,c_{21}\,c^2_{31}\,c_{32}\,
\sin\delta\right]_{\rm m}$
is $\frac{1}{6\sqrt{3}}$. 
If we now employ the elegant three generation relation \cite{HS00}
\beq
\frac{J_{\rm m}}{J_{\rm v}}=\frac
{\left[ \delta m^2_{21} \delta m^2_{32} \delta m^2_{31} \right]_{\rm v}}
{\left[ \delta m^2_{21} \delta m^2_{32} \delta m^2_{31} \right]_{\rm m}}
\label{HSreln}
\eeq
relating vacuum and matter Jarlskog factors 
and mass-squared differences, then the
T-violation asymmetry in matter can be written as
\beq
P^{\not {\rm T}}_{\rm m}= 16 J_{\rm v}
\left[ {(\delta m^2_{21}\delta m^2_{32}\delta m^2_{31})|_{\rm v} }
\over
 {(\delta m^2_{21}\delta m^2_{32}\delta m^2_{31})|_{\rm m} }\right] \,
 \sin\Delta^{\rm m}_{21} \sin\Delta^{\rm m}_{32} \sin\Delta^{\rm m}_{31} \,.
\label{Tmat2}
\eeq
For small distances
such that $\sin \Delta \approx \Delta$ for all 
$\Delta$'s, both in vacuum and matter, then 
$P^{\not {\rm T}}_{\rm v} \approx P^{\not {\rm T}}_{\rm m}$.
That is, at these distances matter effects are negligible.
At longer lengths, the bracketed ratio in eq.(\ref{Tmat2})
can provide an enhancement if
one of the $\delta m^2 |_{\rm m}$ becomes
small compared to its vacuum value,
i.e. near a resonance. 
The first resonance is encountered when
\beq
E_{\nu} \simeq
{ \delta m^2_{21}|_{\rm v} \cos 2\theta^{\rm v}_{21}
\over 
2\sqrt{2} G_F  N_e } \equiv E^{1R}_{\nu} 
\,,
\label{A21}
\eeq
for which
\beq
\delta m^2_{21}|_{\rm m} \simeq \delta m^2_{21}|_{\rm v} 
\sin 2\theta^{\rm v}_{21}
\quad{\rm and}\quad
\delta m^2_{32}|_{\rm m} \simeq \delta m^2_{32}|_{\rm v}\,,
\label{R21}
\eeq
and the bracketed expression in eq.\ (\ref{Tmat2}) becomes 
\beq
{1 \over \sin2\theta^{\rm v}_{21}} \, .
\eeq
The second resonance occurs when
\beq
E_{\nu} \simeq
{ \delta m^2_{32}|_{\rm v} \cos 2\theta^{\rm v}_{31}
\over 
2\sqrt{2} G_F  N_e } \equiv E^{2R}_{\nu} 
\,,
\label{A32}
\eeq
for which
\beq
\delta m^2_{32}|_{\rm m} \simeq \delta m^2_{32}|_{\rm v} 
\sin 2\theta^{\rm v}_{31}
\quad{\rm and}\quad
\delta m^2_{21}|_{\rm m} \simeq \delta m^2_{32}|_{\rm v}\,,
\label{R32}
\eeq
and the bracketed expression in eq. (\ref{Tmat2}) becomes
\beq
{\delta m^2_{21}|_{\rm v} \over \delta m^2_{32}|_{\rm v}
\sin 2\theta^{\rm v}_{31}}.
\eeq

\begin{figure}[t!] 
\includegraphics{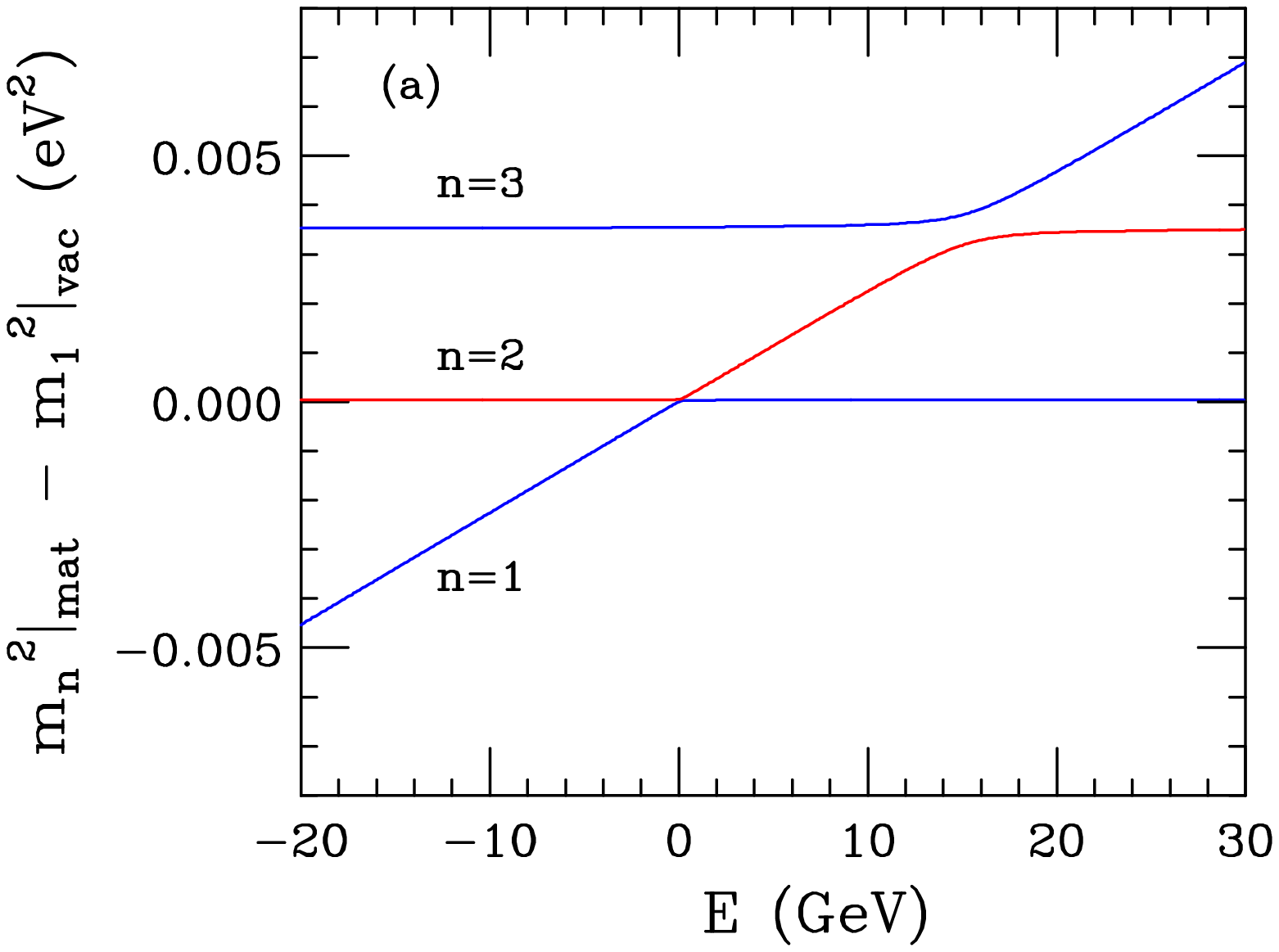}
\includegraphics{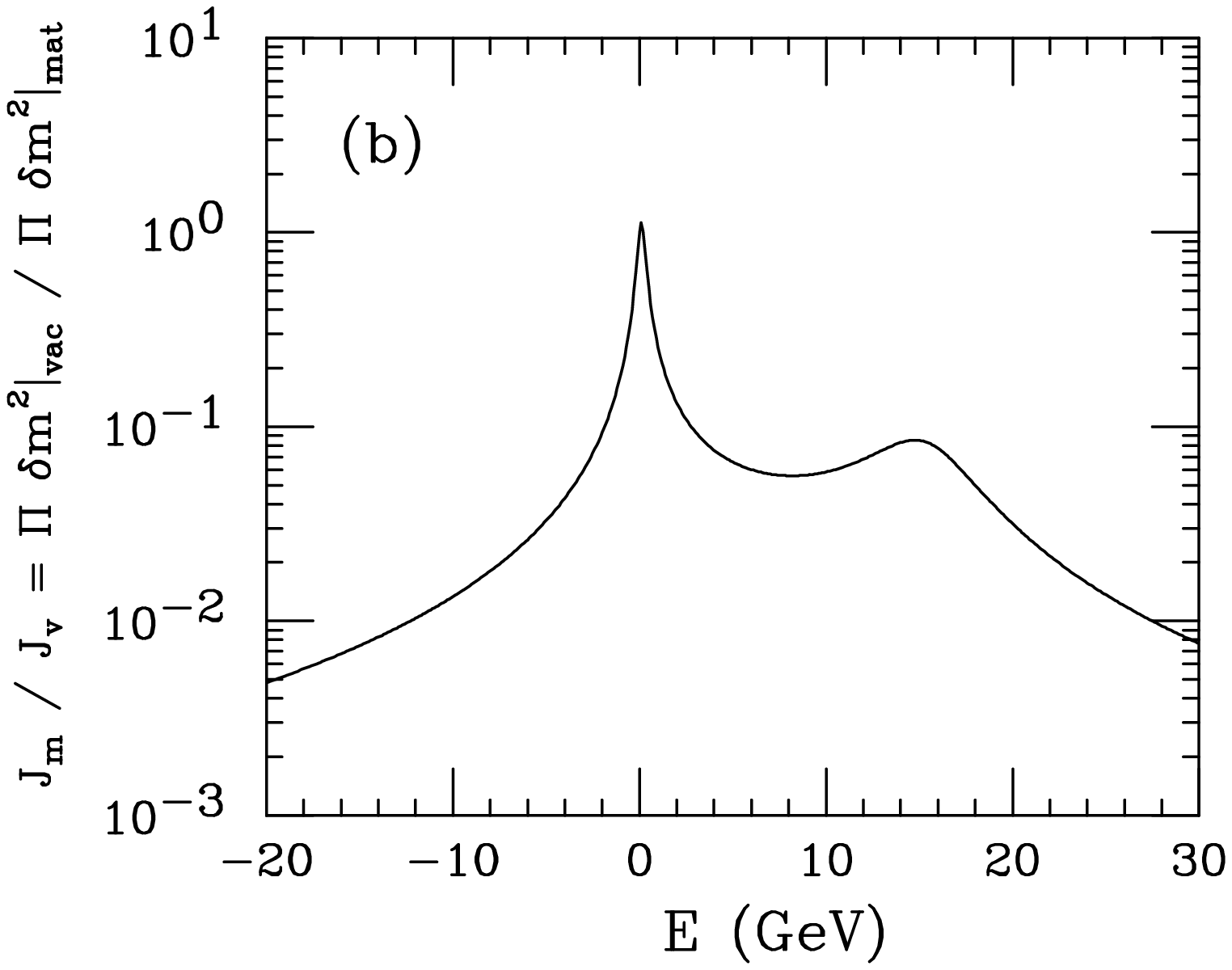}
\vspace{2.25in}
\caption{Shown are 
(a) the neutrino mass-squared eigenvalues in matter
and (b) the ratio $J_{\rm m}/J_{\rm v}$,
for the parameters listed in eq.~(15),
as a function of the neutrino energy.
Positive energies correspond to neutrinos, and 
negative energies correspond to anti-neutrinos
(vice versa for inverted  $\delta m^2$'s).
}
\label{fig:dm2}
\end{figure}

To illustrate how matter affects the Jarlskog factor, $J_{\rm m}$, 
we have chosen representative values for 
both the earth's density and the neutrino vacuum parameters:
a matter density typical of the earth's mantle 
($\sim 3 ~{\rm g ~cm}^{-3}$)
and neutrino masses and mixing
which are centered in the allowed regions for
the atmospheric neutrinos and the large angle MSW solar solution.
The chosen vacuum values are
\begin{eqnarray}
\delta m^2_{32}|_{\rm v} & =  3.5 \times 10^{-3}  {\rm eV}^2  \quad \quad 
\sin^22\theta_{32}^{\rm v} & = 1  \nonumber \\
\delta m^2_{21}|_{\rm v} & =  5.0 \times 10^{-5} {\rm eV}^2 \quad \quad
\sin^22\theta_{21}^{\rm v} & =  0.8  \\
\delta m^2_{31}|_{\rm v} & =  3.5 \times 10^{-3} {\rm eV}^2 \quad \quad
\sin^22\theta_{31}^{\rm v} & =  0.03 \nonumber \\
& \sin \delta^{\rm v}   ~=  ~1 & \nonumber \\
& Y_e ~\rho  ~= ~1.5 ~{\rm g ~cm}^{-3}\, . & \nonumber
\label{pmtrs}
\end{eqnarray}
The CP and T-violating angle $\delta^{\rm v}$ has been chosen to be
maximal, $\pi \over 2$.
The value of $\sin^22\theta_{31}^{\rm v}$
is chosen to be half an order of magnitude below the Chooz limit \cite{CHOOZ}. 
Using these parameter values, the matter mass-eigenvalues
and the ratio $J_{\rm m}/J_{\rm v}$, equal to the 
bracketed ratio of eq.\ (\ref{Tmat2}),
is given in fig. \ref{fig:dm2}. 
Note that away from the resonances,
$J_{\rm m}/J_{\rm v}$ is much less than one, 
suggesting that  matter effects suppress the {\it amplitude} 
of the T-violating oscillation.
In contrast, there are peaks in the ratio $J_{\rm m}/J_{\rm v}$ 
near the resonances.

In principle, the peaks in $J_{\rm m}/J_{\rm v}$ at the resonances
become significant enhancements, $\gg 1$, when either
\beq
\sin 2\theta^{\rm v}_{21} \ll 1\,,
\quad {\rm or} \quad
\sin 2\theta^{\rm v}_{31} \ll {\delta m^2_{21}|_{\rm v} 
\over \delta m^2_{32}|_{\rm v}}
\lsim {\cal O}(10^{-2})\,,
\label{eqn:pks}
\eeq
holds. 
(Neither of these conditions is satisfied in our representative example.) 
However if either of these conditions holds, then the vacuum
Jarlskog factor $J_{\rm v}$ of eq.\ (\ref{Jvac}) is itself very small.
The message here is that the Jarlskog factor 
can be strongly enhanced by matter, as in \cite{YKT00},
only when the vacuum value is very small to begin with;
the enhancement never produces a large value of $J_{\rm m}$.
A quantitative view of the impossibility of matter to produce 
a truly large amplitude results when the explicit expression 
for $J_{\rm v}$ in eq.\ (\ref{Jvac}) is substituted 
into eq.\ (\ref{Tmat2}).
The result is
\beq
P^{\not{\rm T}}_{\rm m}= 
2 \cos \theta^{\rm v}_{31} \, \sin(\delta^{\rm v})
\left[
{[(\sin2\theta_{21}\delta m^2_{21})(\sin2\theta_{32}\delta
m^2_{32})(\sin2\theta_{31}\delta m^2_{31})]_{\rm v}}
\over
{[\delta m^2_{21}\quad\delta m^2_{32}\quad\delta m^2_{31}]_{\rm m} }\right] \,
 \sin\Delta^{\rm m}_{21} \sin\Delta^{\rm m}_{32} \sin\Delta^{\rm m}_{31} \,.
\nonumber
\label{heuristic}
\eeq
As seen from eqs.\ (\ref{R21}) and (\ref{R32}),
at either resonance the bracketed factor in this equation 
does not become large.
What the resonance manages to do is to cancel the small vacuum 
value of $\sin2\theta^{\rm v}_{21}$ or $\sin2\theta^{\rm v}_{31}$
in the {\it amplitude} ($16 J_{\rm v}$) of the T-violating oscillation.
But accompanying even this cancellation is a negative consequence 
for the associated oscillation lengths,
to which we now turn.

\section{Baseline Limitations}
A significant enhancement of T-violating oscillation 
amplitudes requires a small-angle resonance.  
The conditions for this are either
\beq
\delta m^2_{21}|_{\rm m} \ll \delta m^2_{21}|_{\rm v} 
\quad {\rm or} \quad
\delta m^2_{32}|_{\rm m} \ll \delta m^2_{21}|_{\rm v} \,. 
\eeq
These conditions in turn imply that the oscillation length in 
matter associated
with the smallest $\delta m^2$ is larger than the largest
oscillation length in vacuum, the one 
associated with the smallest vacuum $\delta m^2$,
i.e. $\delta m^2_{21}|_{\rm v}$.
Thus, there is the danger that for an enhanced amplitude, 
the baseline requirement will
exceed the capability of a terrestrial experiment.
In fact, this happens.

Assuming hierarchical mass splittings,
the fast oscillation governed by the larger 
\beq
\dplus\equiv {\rm max}
[ \Delta^{\rm m}_{21},\Delta^{\rm m}_{32} ] 
\equiv {\delta m^2_+ L \over 4 E_{\nu}}
\eeq
will be amplitude modulated by the slow oscillation, governed by
\beq
\dminus\equiv {\rm min} 
[ \Delta^{\rm m}_{21},\Delta^{\rm m}_{32} ] 
\equiv {\delta m^2_- L \over 4 E_{\nu}}.
\eeq
The consequence is that the 
T-violating oscillation is bounded above by 
$\sin{(\dminus)}\cos^2{(\dminus/2)}$,
and below by $-\sin{(\dminus)}\sin^2{(\dminus/2)}$.
An idealized experiment would resolve the fast oscillation
and measure the difference of the bounding envelopes.
This difference is just $\sin(\dminus)$, which is maximized at
$(2n+1)\pi/2$.  However, a realistic measurement will average over
the fast oscillation, thereby measuring the mean of the envelopes
given by $\frac{1}{4}\sin(2\dminus)$,\\
which is maximized at
$\dminus= (2n+1)~{\pi \over 4}$.  The choice $n=0$ minimizes 
the source-detector distance, and therefore optimizes the incident flux.
So we are led to consider $\dminus = {\pi \over 4}$ as the condition which 
maximizes the size of the T-violating asymmetry.
The corresponding distance, $L^{\not{\rm T}}_{\rm m}$, is given by
\beq
L^{\not{\rm T}}_{\rm m} = { \pi E_{\nu} 
\over \delta m^2_- } 
\eeq
which at the first or second resonance (eqs.\ (\ref{A21}) and (\ref{R21}),
or (\ref{A32}) and (\ref{R32})) becomes
\beq
L^{\not{\rm T}}_{\rm m}  
= {0.2 D_{\oplus} \over \tan 2 \theta^{\rm v}_{21} }
\quad {\rm or} \quad
{0.2 D_{\oplus} \over \tan 2 \theta^{\rm v}_{31} }
\eeq
when the conversion
\beq
{ \pi \over  2\sqrt{2} G_F  N_e } = 0.2~D_{\oplus} 
\nonumber
\eeq 
is used to relate the matter scale to the earth's diameter,
$D_{\oplus} \sim$ 13,000 km.
This latter conversion holds for the mantle density given in eq.(\ref{pmtrs})
-- the matter density is a factor of 2 larger (smaller) in the earth's 
core (outer crust).
For a significant amplitude enhancement to
occur the corresponding $\theta^{\rm v}$ is necessarily small.
Therefore it follows that when $J_{\rm m}\gg J_{\rm v}$,
then $L^{\not{\rm T}}_{\rm m}$  
approaches or exceeds the diameter of the earth.
This in turn ensures that for terrestrial experiments
with $L\ll L^{\not{\rm T}}_{\rm m}$,
the resonant $\delta m^2_-$ is such that  
$\sin \dminus \sim \dminus \ll 1$ and
that $P^{\not{\rm T}}_{\rm m}$ is of the same order as 
$P^{\not{\rm T}}_{\rm v}$, i.e.
no large enhancement.

\section{Magnitude of the First Peaks}
Two arguments may be made for the importance of the very first,
or first few,  peaks as the target for measurement.
The first argument is that the T-violating oscillation length in the earth,
as discussed above, is characteristically of order of the earth's diameter
or longer.  Thus, even long baseline experiments may be limited to the 
first few peaks.
The second argument is that the $1/L^2$ fall-off of the incident neutrino
flux for any aperture-limited experiment favors shorter distances.\footnote
{When $\sin(\dminus)\sim \dminus$ applies, one power of $L^{-1}$
is compensated.}

In vacuum, the first few peaks in the T-violating asymmetry given by
eq.\ (\ref{Tvac}) occur when\\
$\sin\Delta^{\rm v}_{21} \sin\Delta^{\rm v}_{32} \sin\Delta^{\rm v}_{31}
\approx { \delta m^2_{21}|_{\rm v} \over \delta m^2_{32}|_{\rm v} }
\Delta^{\rm v}_{32}\sin^2 \Delta^{\rm v}_{32}$ is maximized.
The rough location of these peaks is given by
\beq
{L_{\rm v}\over E_{\nu}} \sim (2n+1)\,2\pi
~{1 \over \delta m^2_{32}|_{\rm v}}\,,
\label{LonE}
\eeq
for n=0,1,2~$\cdots$ until the approximation 
$\sin \Delta^{\rm v}_{21} \approx \Delta^{\rm v}_{21}$ is no longer valid.
The asymmetry at the n-th peak is equal to
\beq
(2n+1)\,8\pi
~J_{\rm v} ~{ \delta m^2_{21}|_{\rm v} \over \delta m^2_{32}|_{\rm v} }\,.
\label{asymsize}
\eeq
Thus the asymmetry and the distance to the n-th peak both grow as (2n+1).
In particular the second peak is three times larger than the first peak
and three times further out.
Remember however that the neutrino beam intensity falls as 
$L^{-2}$ for fixed $E_\nu$, disfavoring the more distant peaks. 

For the first peak, a numerical calculation produces an estimate
more accurate than eqs.\ (\ref{LonE}) and (\ref{asymsize}); 
the result is
\beq
\Delta^{\rm v}_{32}
\sim {7 \pi \over 12}\,,
\quad {\rm i.e.} \quad 
{L_{\rm v}\over E_{\nu}} \sim {7 \pi \over 3} 
~{1 \over \delta m^2_{32}|_{\rm v}}\, ,
\label{optvac}
\eeq
with the size of the asymmetry at the first peak being
\beq
\sim ~{8.7  ~\pi }
~J_{\rm v} ~{ \delta m^2_{21}|_{\rm v} \over \delta m^2_{32}|_{\rm v} }\,.
\label{asym1}
\eeq
So even the first peak is reasonably approximated by the general results
above.

In matter, the physics is more complex since the $\delta m^2$'s 
change with energy. 
We start by looking at the product of the three sine terms
\beq
\sin(\dminus)\sin(\dplus)\sin(\dminus+\dplus)\,.
\nonumber
\eeq
The first peak of this  product occurs when 
\beq
\dminus+\dplus \sim {7 \pi \over 12}
\quad {\rm if} \quad \dminus \ll \dplus\,,
\eeq
and at 
\beq
\dminus +\dplus = {2 \pi \over 3}
\quad {\rm if} \quad \dminus = \dplus.
\eeq  
In terms of $(\dminus+\dplus)$, the first peak moves monotonically
from $\sim {7 \pi \over 12}$ to $2 \pi \over 3$
as the ratio of $\dminus$ to $\dplus$ changes from 0 to 1.
At the first peak this product of the sines may be written as 
\beq
\eta~
{ 2 \dminus \dplus 
\over
(\dminus +\dplus)^2 }
\quad = \quad 
\eta~
{ 2 \delta m^2_- \delta m^2_+
\over
(\delta m^2_- + \delta m^2_+)^2 }
\eeq
where $\eta$ is slowly varying, monotonically increasing, function of 
$\frac{\dminus}{\dplus} = \frac{\delta m^2_-}{\delta m^2_+}$ with
$\eta(0) \approx 0.86$ and $\eta(1) = 3\sqrt{3}/4 \approx 1.30$.

Thus the full T-violating asymmetry, eq.(\ref{Tmat2}), 
at the first peak is
\beq
P^{\not {\rm T}}_{\rm m}= 32 ~J_{\rm v}
~\eta 
\left[ { (\delta m^2_{21}\delta m^2_{32}\delta m^2_{31})|_{\rm v}
\over
 (\delta m^2_- + \delta m^2_+)^3 } \right] \, .
\label{Tmat3}
\eeq
For energies between the two resonances, eqs.(\ref{A21}) and (\ref{A32}),
the following sum-rule holds 
\beq
(\delta m^2_- + \delta m^2_+) 
\approx \delta m^2_{32}|_{\rm v}\,,
\label{Msqid}
\eeq
as evidenced in Fig. \ref{fig:dm2} (a).
This is a good approximation provided both
$\delta m^2_{21}|_{\rm v} \sin 2\theta^{\rm v}_{21}$
and
$\delta m^2_{32}|_{\rm v} \sin 2\theta^{\rm v}_{31}$
are much smaller than 
$\delta m^2_{32}|_{\rm v}$.

\begin{figure}[t!] 
\includegraphics{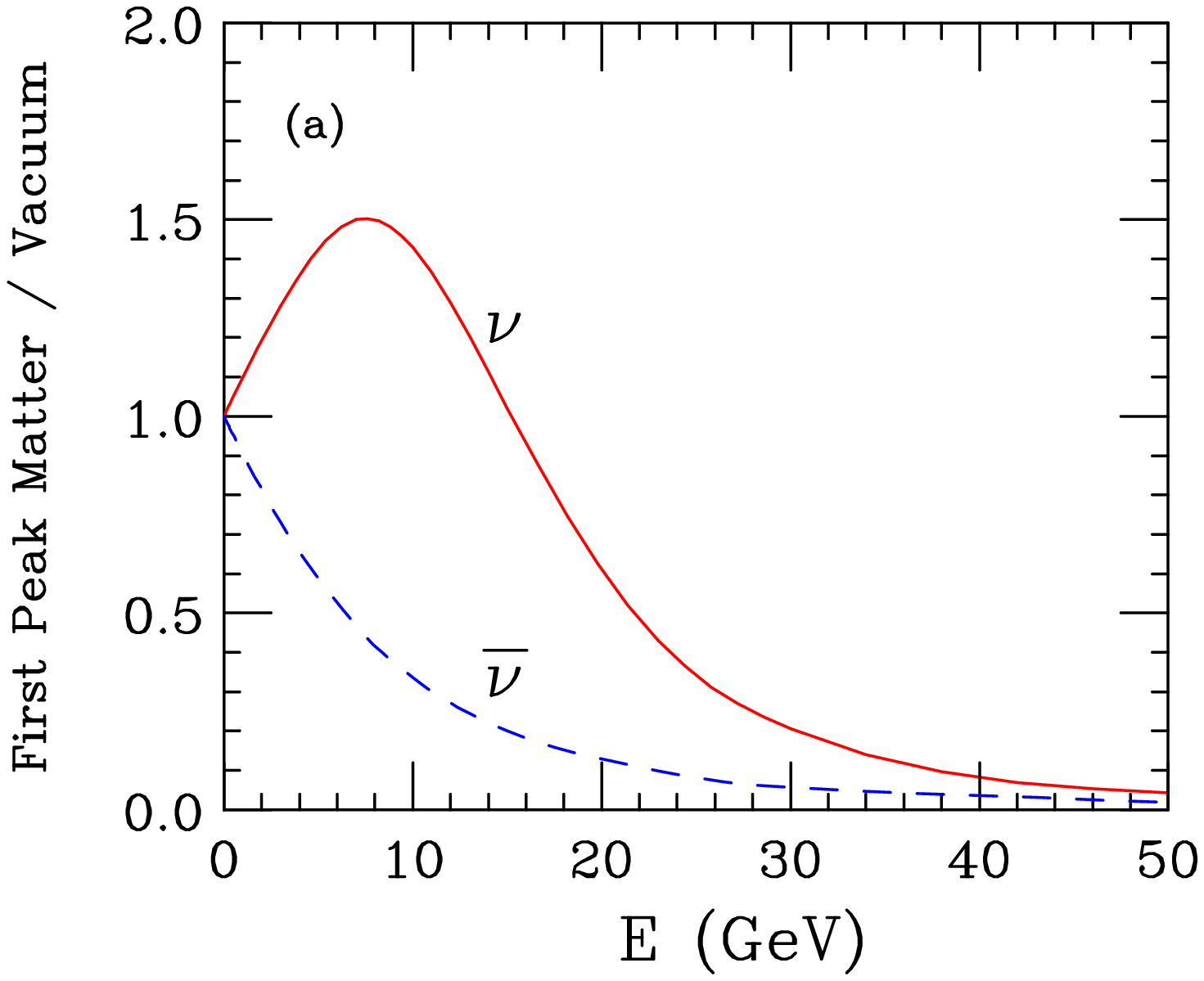}
\includegraphics{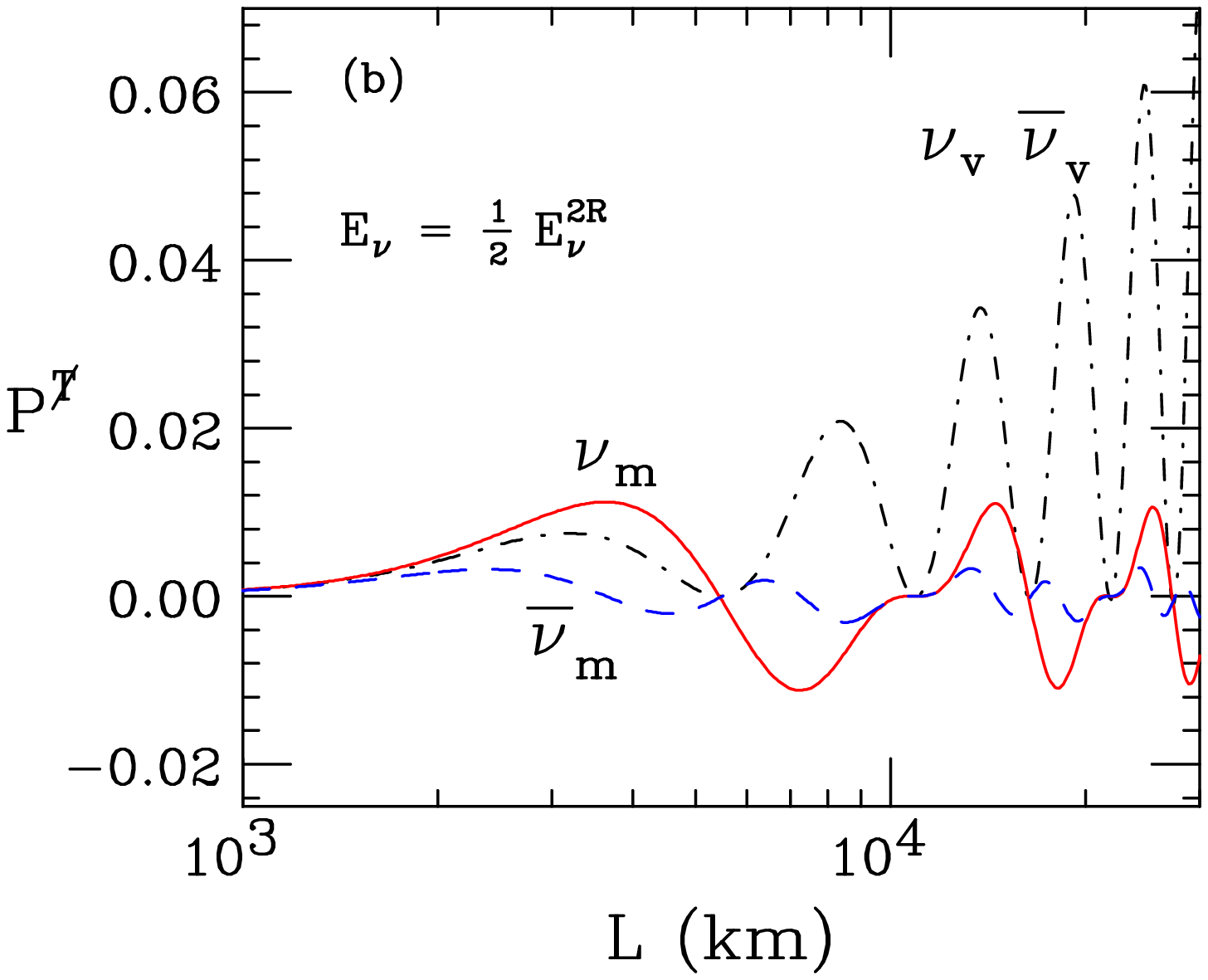}
\vspace{2.25in}
\caption{
Using the parameters given in eq.(15); 
Shown are 
(a) the ratio of the value of the asymmetry at the 
first peak in matter over vacuum as function of the neutrino energy --
for neutrinos the ratio peaks at half the second resonant energy.
(b) the asymmetry $P^{\not {\rm T}}$ versus distance for neutrinos and
anti-neutrinos in matter and in vacuum for an energy (7.6 GeV) equal to
half of the second resonant value, as specified in eq.\ (\ref{halfER}).}
\label{fig:peaks}
\end{figure}

Below the first resonance and above the second resonance,
$(\delta m^2_- + \delta m^2_+)$ grows approximately linearly
with energy from the minimum value,
$\delta m^2_{32}|_{\rm v}$.
Since the function $\eta$ varies little 
it cannot compensate for the increase in size of
$(\delta m^2_- + \delta m^2_+)$ for energies below the first resonance or
above the second resonance.
Therefore the maximum value of the first peak of the T-violating
asymmetry occurs between the resonances where $\eta$ is maximized, i.e.
\beq
\delta m^2_- = \delta m^2_+ \approx \half \delta m^2_{32}|_{\rm v} \,.
\eeq
This is shown in Fig. \ref{fig:peaks}(a) where
the ratio of
the asymmetry in matter and in vacuum at their respective first peaks,
is plotted versus energy. (The vacuum value of the first peak is
energy independent, eq.(\ref{asym1})).

The neutrino energy for the maximum first peak in matter
is given by (recall eq.\ (\ref{A32}))
\begin{eqnarray} 
E_{\nu} \simeq
\half E^{2R}_{\nu} & = & { \delta m^2_{32}|_{\rm v} \cos 2 \theta_{31}^{\rm v}
\over 
4 \sqrt{2} G_F  N_e  } \\
& \sim & 7.6 \cos 2 \theta_{31}^{\rm v}           
~\left[ {\delta m^2_{32}|_{\rm v} \over 3.5\times 10^{-3}~{\rm eV}^2}\right]
~\left[ { 1.5 ~{\rm g ~cm}^{-3} \over Y_e \rho}\right] ~{\rm GeV}\, ,
\label{halfER}
\end{eqnarray}
and the peak occurs at a distance fixed by $\dplus=\dminus=\pi/3$, i.e.
\beq
L=\frac{8\pi}{3}\frac{E_{\nu}}{\delta m^2_{32}|_{\rm v}} \,
=~\frac{4\pi}{3}{ \cos 2 \theta_{31}^{\rm v}
\over 
2 \sqrt{2} G_F  N_e  }
\sim 3600\;\cos 2 \theta_{31}^{\rm v}
~\left[ { 1.5 ~{\rm g ~cm}^{-3} \over Y_e \rho}\right] ~{\rm km} \, .
\label{Lspecial}
\eeq
Note that this distance is independent of $\delta m^2|_{\rm v}$.
The value of the T-violating asymmetry at this peak is
\beq
{ 24 \sqrt{3} } J_{\rm v} 
{\delta m^2_{21}|_{\rm v} \over \delta m^2_{32}|_{\rm v} }\,.
\eeq

The matter asymmetry at this peak is 52\% larger than the 
asymmetry at the first peak in vacuum, 
given in eq.\ (\ref{asym1}).
From Fig. \ref{fig:dm2} (b) we see that, for our chosen parameters,
matter effects at $E_\nu = 7.6$~GeV actually suppress $J$
by more than an order of magnitude
which is more than compensated by an increase in the 
product of sines due to matter.
But at this energy the $L$ value for the
first peak in matter exceeds the same in vacuum by 14\%.

In Fig.~\ref{fig:peaks}(b)  we show the repeating (in $L$)
peaks for the T-violating asymmetry with
the energy fixed at one half the second resonant value
for both neutrinos and anti-neutrinos in matter and vacuum.
For neutrinos this energy gives the largest first peak 
in matter and
all subsequent peaks have the same magnitude but alternate in sign.

To go beyond the first peak in the asymmetry is quite complicated.
For the second peak, 
the general features are that matter effects, if important,
suppress the magnitude of the asymmetry compared to vacuum values;
also, these effects can flip the sign of the asymmetry
compared to the first peak. 
There are two situations which allow a simple complete analysis.
The first situation results when the neutrino energy
is chosen so there is a substantial mass hierarchy
(i.e. near one of the resonant energies).
The size and position of the peaks in this situation 
closely follow the vacuum
case discussed earlier, see eqs.\ (\ref{LonE}-\ref{asym1}).
For the neutrino parameters used in this paper,
this situation occurs only at low neutrino energy $<$ 3 GeV or 
if $\theta^{\rm v}_{31}$ is small near
$E_{\nu} = E^{2R}_\nu \sim 15$~GeV.
However, at the higher-energy resonance, 
only that part of the beam which is
within a few GeV of the resonance energy, 
so as to maintain the extreme $\delta m^2 |_{\rm m}$ hierarchy,  
contributes to an unsuppressed asymmetry.
The second situation results when the neutrino energy and travel length
are such that the product of the three sines is at its maximize value.
This case is solved in generality in the Appendix and is relevant for
case when $E_{\nu} = \half E^{2R}_\nu \sim 7.6$~ GeV discussed above.

\section{Summary and Conclusions}
Even though matter effects can significantly enhance the Jarlskog factor
in cases where the vacuum value is small,
this enhancement does not lead to
large enhancements of the T-violating probabilities for terrestrial
experiments.  
The reason for this is that associated with this enhancement of the Jarlskog
factor is an increase in the longest oscillation length so that
for the neutrino parameters suggested by current experiments 
with small vacuum Jarlskog factor, the enhancement
occurs for distances beyond the earth's diameter.

However we have shown that the first peak in the T-violating probability
can be enhanced in matter as a result of
an enhancement of the oscillating factors,
which more than compensates for the suppression of the Jarlskog factor.
This first peak is experimentally the most accessible. 
The enhancement of the first peak occurs for neutrino energies between the 
two resonant energies and has a broad maximum midway between these
two resonant energies that is 50\% larger than the
first peak in vacuum.
The enhanced first peak occurs for a neutrino
travel distance of
$2\pi\cos 2 \theta_{31}^{\rm v}/ 3 \sqrt{2} G_F  N_e$
which is $\sim$3600 km.
Note that this distance depends only on the density of the earth
and not on the parameters of the neutrinos
(assuming $\cos 2 \theta_{31}^{\rm v} \sim 1$).

Away from the resonant energies the second and higher peaks
in the T-violating probability are generally suppressed
compared to their growing vacuum counter parts.
For neutrinos with energies higher than the higher resonant energy
and for anti-neutrinos of all energies the first peak in matter is suppressed
compared to the corresponding peak in vacuum.
Thus the optimum selectable parameters for the observation of
T-violation are a neutrino energy midway between the resonance energies
and a travel length of $\sim$3600 km.  

Application of these ideas to the experimentally more accessible
CP-violation using neutrino factory beams is under investigation.

\section*{Acknowledgements}
We thank the Aspen Center for Physics for providing a
stimulating working environment were this work began.
SP would like to thank Boris Kayser for useful discussions.
This work was supported in part by the
DOE grant no.\ DE-FG05-85ER40226, and
Fermilab is operated by URA under DOE contract No.~DE-AC02-76CH03000.

%

\section{APPENDIX: Solutions for Maximum Peaks}
In this appendix we derive the energy, distance, and peak height
associated with each maximum product 
$\sin(\dminus)\sin(\dplus)\sin(\dplus+\dminus)$.
Above $\sim 2$~GeV, safely away from the lowest-energy resonance,
the prefactor $J_{\rm m}/J_{\rm v}$,
exhibited in Fig.\ 1b, is a relatively 
slowly-varying function of $E$.
Thus, we expect the conditions for the 
peaks in the asymmetry to be well-approximated by the  conditions
for the maxima in the product of sines.  
The error inherent in this approximation is almost certainly
less than the observational error in the neutrino energy inferred
from the charged-current measurement.

The largest value of the product 
$\sin(\dminus)\sin(\dplus)\sin(\dminus+\dplus)$
is $(\sqrt{3}/2)^3$ when the energy of the neutrinos 
and the detector distance $L$ are such that two 
conditions on the phases of 
the slow and fast oscillations are met.  
The conditions are
\beq
\Delta_{\pm}=\frac{\pi}{3}\,n_{\pm}\,,
\label{twoconds}
\eeq
where
\beq
n_- \quad {\rm and} \quad n_+ =1\;({\rm mod}\,3) 
\quad {\rm or} \quad 2\;({\rm mod}\,3)\,,
\label{ndefined}
\eeq
and $n_+ \ge n_-$ by definition.
One of the conditions in eq.\ (\ref{twoconds}) may be replaced 
by an equivalent commensurability condition,
\beq
\frac{\dminus}{\dplus}
=\frac{\delta m^2_-}{\delta m^2_+}
=\frac{n_-}{n_+}\,.
\label{commens}
\eeq

With two variables ($E$ and $L$) under experimental control,
the two independent conditions of 
eqs.\ (\ref{twoconds}--\ref{commens})
can always be satisfied by a designer experiment, in principle.
The commensurability condition (\ref{commens}) is satisfied by the choice
of neutrino energy.
Reference to Fig.\ 1a reveals that for $E$ between the two resonant energies,
a linear relation holds:
\beq
E\approx \frac{\delta m^2_\mp}{\delta m^2_{32}|_{\rm v}}\,E_R\,,
\label{linearE}
\eeq
where here and in what follows the $\mp$ holds for 
$E\stackrel{<}{>}\half E_R$.%
This equation leads to the desired constraint on the requisite energy:
\beq
E=\frac{\Delta^{\rm m}_\mp}{\dminus +\dplus}\;E_R
=\frac{n_\mp}{n_- +n_+}\;E_R\,.
\label{Especial}
\eeq
The requisite length is obtained from
$L=4E\Delta_{\pm}/\dms_{\pm}$.
Substituting for $E$ from eq.\ (\ref{linearE}) and for $\Delta^{\rm m}_\mp$
from eq.\ (\ref{twoconds}),
one gets
\beq
L=\frac{4\pi}{3}\frac{E_R}{\delta m^2_{32}|_{\rm v}}\;n_\mp \,.
\label{Lspecial2}
\eeq

To summarize these results, 
the T-violating oscillation asymmetry
is maximized when the neutrino energy satisfies eq.\ (\ref{Especial}),
and the length satisfies eq.\ (\ref{Lspecial2}).
At fixed optimizing energy $E$, i.e. fixed $n_+/n_-$,
eq.\ (\ref{Lspecial2}) predicts recurring peaks at distances 
related to the distance to the first peak
\beq
L_1=\frac{4\pi}{3}\frac{E_R}{\delta m^2_{32}|_{\rm v}}
\sim 3600\;\cos 2 \theta_{31}
~\left[ { 1.5 ~{\rm g ~cm}^{-3} \over Y_e \rho}\right] ~{\rm km}
\label{L1}
\eeq
by $L= L_1 \times n_\mp$,
with $n_\mp$ any positive integer allowed by eqs.\ (\ref{Especial}) 
and (\ref{ndefined})
(which excludes multiples of three, at a minimum).

Plotted against distance at fixed $E$, each peak will exhibit 
the same height since the prefactor $J_{\rm m}$ depends on
energy but not distance.
One may ask what energy
optimizes the peak heights.
The prefactor $J_{\rm m}$ is maximized when
$\frac{\delta m^2_+}{\delta m^2_-}$ is maximized.
From eqs.\ (\ref{commens}) and (\ref{Msqid}),
one finds
\beq
\delta m^2_\mp = \frac{n_\mp}{n_- +n_+}\,\delta m^2_{32}|_{\rm v}\,,
\eeq
which, when inserted into eq.\ (\ref{HSreln}), yields
\beq
\frac{J_{\rm m}}{J_{\rm v}}=
\frac{(n_+ +n_-)^2}{n_+ n_-}\;
\frac{\delta m^2_{21}|_{\rm v}}{\delta m^2_{32}|_{\rm v}}\, ;
\eeq
and so the heights of the asymmetry peaks are
\beq
6\sqrt{3} \;
\frac{(n_+ +n_-)^2}{n_+ n_-} \;
\frac{\delta m^2_{21}|_{\rm v}}{\delta m^2_{32}|_{\rm v}}\,.
\eeq
In terms of $E$, given in eq.\ (\ref{linearE}), 
these asymmetry heights are
\beq
6\sqrt{3} \;
\frac{E_R^2}{E(E_R-E)} \;
\frac{\delta m^2_{21}|_{\rm v}}{\delta m^2_{32}|_{\rm v}}\,.
\eeq
As $E$ approaches the low-energy resonant value near zero
or the high-energy resonant value $E_R$,
the assumption that the $\delta m^2$'s vary linearly with energy fails
and our formulae here become invalid.

\end{document}